\documentclass[aps,twocolumn,superscriptaddress, longbibliography]{revtex4-2}
\usepackage{graphicx}
\usepackage{bm}
\usepackage{braket}
\usepackage{pifont}
\usepackage{amsmath}
\usepackage{amsfonts}
\usepackage{amssymb}
\usepackage{graphicx}
\usepackage{color,xcolor}
\usepackage{tensor}
\usepackage[caption=false]{subfig}
\usepackage{natbib}
\usepackage{ulem}
\usepackage{hyperref}
\usepackage{enumitem}
\hypersetup{
    colorlinks=true,
    linkcolor=blue,
    filecolor=black,   
    citecolor=blue, 
    urlcolor=blue,
   	pdftitle={Distributing entanglement in first generation discrete and continuous variable quantum repeaters},
    bookmarks=true,
}
\newcommand{\phantomsubfloat}[1]{
    {
        \captionsetup[subfigure]{labelformat=empty}
        \subfloat[][]{#1}
    }%
}

\begin{document}
\title{Distributing entanglement in first generation discrete and continuous variable quantum repeaters} 
\author{Josephine Dias}
\email{josephine.dias@oist.jp}
\affiliation{Quantum Information Science and Technology Unit, Okinawa Institute of Science and Technology Graduate University, Onna-son$,$ Okinawa 904-0495$,$ Japan}
\affiliation{National Institute of Informatics$,$ 2-1-2 Hitotsubashi$,$ Chiyoda$,$ Tokyo 101-0003$,$ Japan.}
\author{Matthew S. Winnel}
\affiliation{Centre for Quantum Computation and Communication Technology$,$ School of Mathematics and Physics$,$ University of Queensland$,$ Brisbane$,$ Queensland 4072$,$ Australia.} 
\author{William J. Munro}
\affiliation{NTT Basic Research Laboratories $\&$ NTT Research Center for Theoretical Quantum Physics$,$
NTT Corporation$,$ 3-1 Morinosato-Wakamiya$,$ Atsugi$,$ Kanagawa$,$ 243-0198$,$ Japan.}
\affiliation{National Institute of Informatics$,$ 2-1-2 Hitotsubashi$,$ Chiyoda$,$ Tokyo 101-0003$,$ Japan.}
\author{T.C. Ralph}
\affiliation{Centre for Quantum Computation and Communication Technology$,$ School of Mathematics and Physics$,$ University of Queensland$,$ Brisbane$,$ Queensland 4072$,$ Australia.} 
\author{Kae Nemoto}
\affiliation{Quantum Information Science and Technology Unit, Okinawa Institute of Science and Technology Graduate University, Onna-son$,$ Okinawa 904-0495$,$ Japan}
\affiliation{National Institute of Informatics$,$ 2-1-2 Hitotsubashi$,$ Chiyoda$,$ Tokyo 101-0003$,$ Japan.}
\date{\today}
\begin{abstract}
Quantum repeaters are used to overcome the exponential photon loss scaling that quantum states acquire as they are transmitted over long distances. While repeaters for discrete variable encodings of quantum information have existed for some time, novel approaches for continuous variable encoding quantum repeaters have only recently been proposed. In this work, we present a method of using a discrete variable repeater protocol to distribute continuous variable states and utilize it to compare the rates of continuous variable entanglement distribution between first generation continuous and discrete variable quantum repeaters. Such a comparison allows us to begin to benchmark the two quite different approaches.
\end{abstract}
\maketitle

\section{Introduction}
The development of technologies according to the principles of quantum mechanics allows promising real world applications including secure communication \cite{gisin2002quantum, scarani2009security} and quantum information transfer \cite{pirandola2015advances}. However, utilizing these technologies over long distances remains challenging due to fiber loss or free space attenuation \cite{sangouard2011quantum}. A proposed method for allowing the long distance distribution of quantum states are quantum repeaters \cite{briegel1998quantum}. In this model, the long distance is segmented into smaller, more manageable attenuation lengths. Entanglement is distributed along these lengths followed by nested entanglement purification \cite{bennett1996purification,deutsch1996quantum,pan2001entanglement,kwiat2001experimental,pan2003experimental,yamamoto2003experimental,duer2007entanglement} and swapping \cite{ziukowski1993event,bose1998multiparticle,pan1998experimental,riedmatten2005long}. By using a quantum repeater, the exponential error probability scaling with distance that would arise from direct transmission, can be overcome \cite{duan2001long} regardless of whether discrete or continuous variable encoded information is being sent. 

Quantum repeater protocols have existed for discrete-variable (DV) encodings since the mid nineties \cite{briegel1998quantum}, and have been through various iterations of protocol improvements since then \cite{munro2015inside, sangouard2011quantum}. The evolution of quantum repeater protocols has been broadly categorized into three distinct generations \cite{munro2015inside,muralidharan2016optimal}. First generation repeaters are characterized by their use of heralded entanglement generation between repeater nodes, and nested entanglement purification protocols \cite{briegel1998quantum, duer1999quantum}. While first generation repeater protocols were limited due to the time associated with two-way communication of successful generation and purification they were also much easier to implement, given the state of the current technology. However, second \cite{jiang2009quantum, munro2010quantum, zwerger2014hybrid}  and third \cite{munro2012quantum, fowler2010surface, muralidharan2014ultrafast, azuma2015all} generation protocols utilize quantum error correction to be much more efficient and can achieve much higher communication rates. However, all protocols have mostly been focused on sending discrete variable information. Continuous variables provide an alternate approach to encoding information, but have also suffered from the inability to transmit them over long distances. 

 Recently, there have been several proposals for repeater protocols that work with continuous variable (CV) encodings  for the first time \cite{dias2017quantum, furrer2018repeaters, seshadreesan2020continuous,dias2020quantuma, ghalaii2020capacity, winnel2021overcoming}. Such approaches are at the earliest stage of their development but share many similarities to the initial first generation DV schemes \cite{briegel1998quantum,duer1999quantum,duan2001long}. The DV and CV regimes of quantum information are subject to their own unique advantages and disadvantages. For example, CV encodings offer in-principle easier state generation, manipulation and detection \cite{braunstein2005quantum} and also the possibility of more compatibility with existing telecommunications infrastructure \cite{kumar2015coexistence}. In the DV regime maximally entangled states are a physically realizable resource, however the CV counterpart is unphysical and requires infinite energy \cite{giedke2002characterization}. In considering long distance communication, loss has the affect of adding noise to CV states,  which is not the case for DV encodings though loss does affect the probability of successfully establishing entangled links between adjacent nodes \cite{scarani2009security}. It is of significant interest therefore to determine how these first generation continuous variable quantum repeaters perform compared to the existing DV counterparts. 

\begin{figure}
\centering
\includegraphics[width=0.97\linewidth]{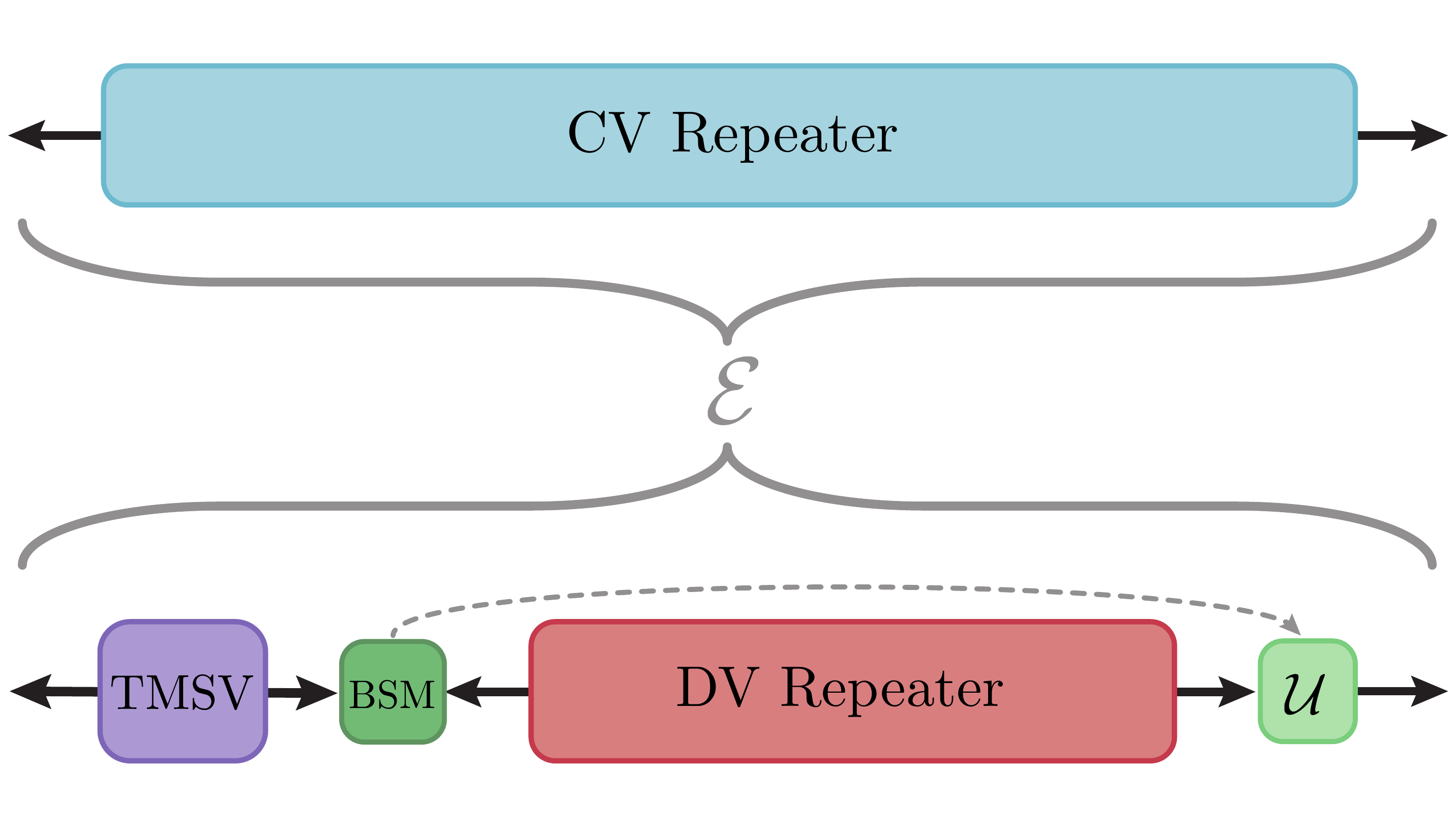}
\caption{Illustration of how we compare rates of two mode CV entanglement distribution between CV and DV repeater protocols. DV repeaters may be utilized to distribute CV entangled states via multi-mode discretized teleportation \cite{andersen2013high}, we have shown here this teleporation protocol operating with a single mode. In our comparison we ensure that the distributed states have the same amount of entanglement.}
\label{fig:RepeaterComparison}
\end{figure}

This work aims to answer this question by performing a comparison between the first generation CV and DV repeaters. In this comparison, we will compare how efficiently both CV and DV repeaters can distribute CV entangled resource states. 
  The performance metric we will use is the \textit{repeater rate} \(R_{\text{rep}}\) which is the rate of generation of entangled states, in units of entangled pairs per second. To ensure a fair comparison we will compare the repeater rates of both CV and DV schemes distributing two mode squeezed states of a similar entanglement level. Both repeaters will be modeled distributing a two-mode squeezed vacuum (TMSV) state (otherwise known as an Einstein-Podolsky-Rosen pair \cite{einstein1935can, weedbrook2012gaussian}) of form
\begin{equation}
\ket{\chi}= \sqrt{1-\chi^2} \sum_{n=0}^\infty \chi^n\ket{n} \ket{n} ,
\label{eq:EPR}
\end{equation}
in the Fock photon-number basis, where \(\chi\) controls or determines the strength of the squeezing. The mean photon number of one mode of this state is \(\bar{n}={\chi^2}/\left({1-\chi^2}\right)\). A basic conceptual diagram of the comparison is shown in Figure~\ref{fig:RepeaterComparison}. While the CV repeater may be used to distribute CV entangled states, in order to distribute these states via DV repeaters, we use a CV teleportation technique \cite{andersen2013high} discussed in more detail in Sec.~\ref{subsec:DV}. The essentials of our comparison are the following: (i) both protocols aim to distribute an equal amount of CV entanglement (TMSV state); (ii) the DV protocol uses qubit distillation, entanglement swapping and teleportation techniques; (iii) the CV protocol uses CV entanglement swapping swapping and NLA.
This paper is arranged in the following way: in Section~\ref{sec:DV} we review the discrete variable repeater protocol illustrating how it can be used to distribute continuous variable quantum states. Then in Section~\ref{sec:CV} we review the continuous variable repeater protocol from Ref.~\cite{dias2020quantuma}, as this is the protocol our rate comparison focuses on. We discuss specifics of the rate comparison and present results in Section~\ref{sec:rate} before we summarize and conclude in Section~\ref{sec:conc}.

\section{Discrete variable repeaters \label{sec:DV}}
\begin{figure}
	\centering
	\includegraphics[width=0.9\linewidth]{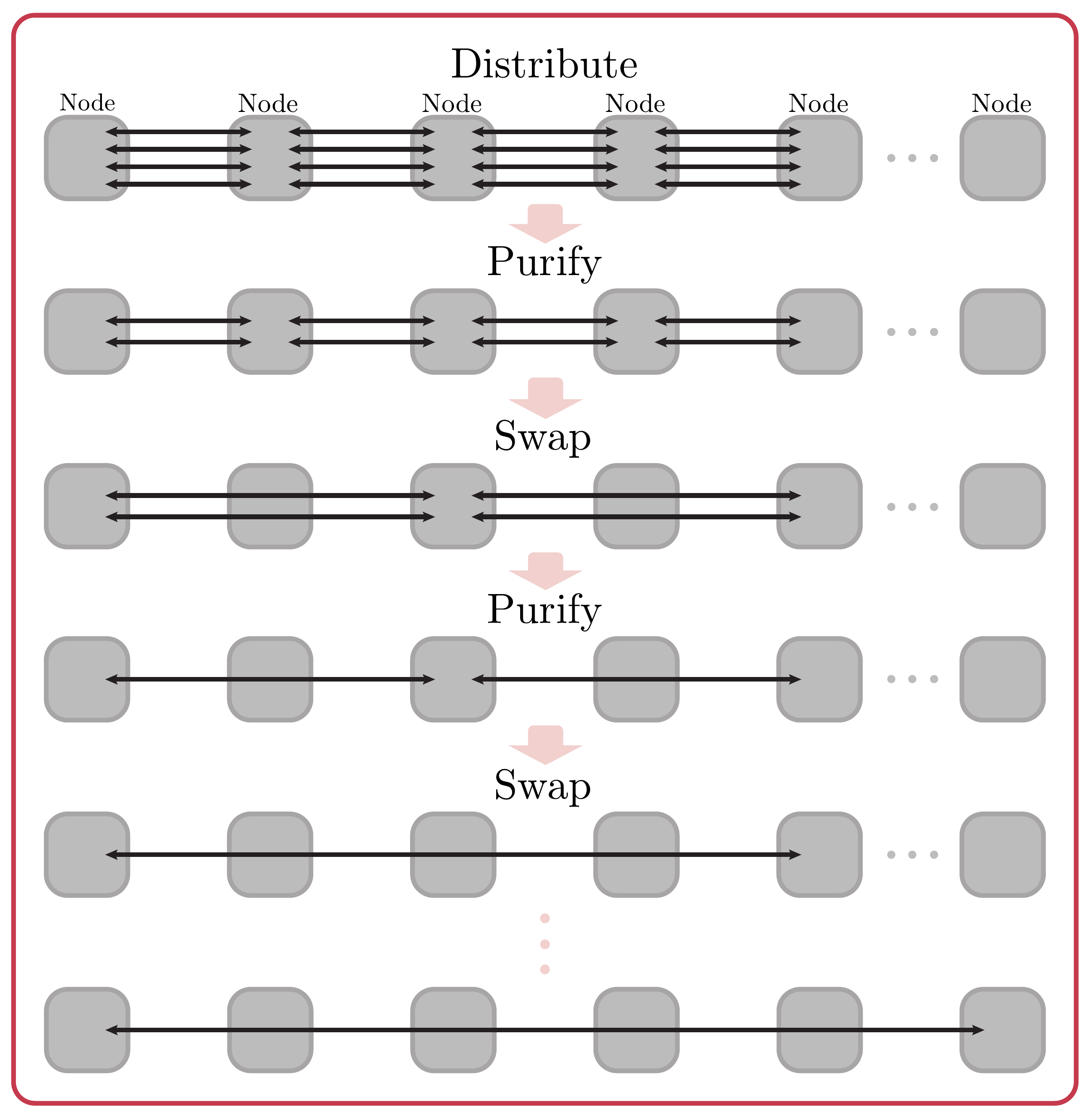}
	\caption{The quantum repeater operation for discrete variables. This illustration shows how the repeater protocol from Ref.~\cite{briegel1998quantum} may operate. Beginning with the initial entanglement distribution of many entangled pairs between each node, subsequent rounds of entanglement purification and swapping follow resulting in a single entangled pair held between both ends of a long distance channel.}
	\label{fig:DVRepeater}
\end{figure}
The 1998 paper by \textit{Briegel et al.} \cite{briegel1998quantum} introduced the concept of a quantum repeater with the goal of overcoming the exponential loss scaling and creating an entangled pair over arbitrary large distances. The Briegel model, now known as a first generation quantum repeater, consists of three core elements: entanglement distribution, entanglement swapping \cite{ziukowski1993event} and nested entanglement purification protocols \cite{bennett1996purification}. It in principle allows one to generate a high quality entangled state between two remote parties, involving many quantum repeaters connecting them, with a polynomial scaling in resources - even when exponential channel loss and imperfect local gates are included. The protocol begins by the distribution of a number of entangled pairs between adjacent nodes in the repeater network. Ideal operation of the repeater would achieve distribution of perfect Bell pairs between nodes of the form,
\begin{equation}
\ket{\Phi^+} = \frac{1}{\sqrt{2}} \left( \ket{01} \ket{10}+\ket{10}\ket{01}\right) , 
\end{equation}
where we have used the dual-rail encoding with $\ket{01}$ and $\ket{01}$ being orthogonal basis vectors spanning the two-dimensional Hilbert space of the qubit. This pair would then be used for subsequent rounds of entanglement swapping until a final pair is produced between both ends of the channel. This ideal situation is unrealistic as we are using the pair source to generate our entangled state. There will be both the vacuum state and higher order photon numbers that contribute to our resulting state. Photon loss in the channel will cause our entangled state to become mixed. Further, local gate operations within the repeater nodes themselves will cause errors. A convenient model for the errors induced by imperfect production is the Werner state \cite{werner1989quantum}:
\begin{align}
{{\rho}}_w&= \frac{4F-1}{3} \ket{\Phi^+}\bra{\Phi^+} + \frac{1-F}{3} \mathrm{I}_4
\label{eq:Werner}
\end{align}
where $\mathrm{I}_4$ is the identity matrix of the two-qubit Hilbert space.  The state \eqref{eq:Werner} has fidelity \(F\) for the required pair \(\ket{\Phi^+}\) but also contains a mixture of all the other Bell states. We acknowledge this is a crude approximation to the real entangled state that will be generated, but serves the purpose to illustrate that the resulting entangled state distributed will not be fidelity one. With two pairs (labeled \({\rho_w}_{12}\),\({\rho_w}_{34}\)) distributed between three nodes, entanglement swapping \cite{ziukowski1993event} proceeds as follows: a local joint Bell-state measurement is conducted between qubits \(2\) and \(3\). The results of that measurement are then sent via a classical communication channel to qubit \(4\) where a Pauli correction is made on qubit \(4\) based on the outcome of the measurement. The result being, a single entangled pair is now shared between the outer nodes \({\rho_w}_{14}\). Beginning with two Werner pairs, each of fidelity \(F\), the fidelity of the swapped pair is given by \cite{munro2015inside}:
\begin{equation}
F_{swap} = F^2 + \frac{\left(1-F\right)^2}{3}  ,
\label{eq:Fswap}
\end{equation}
which is always less than the fidelity of the initial pair \(F\). Further, as a linear optical Bell-state measurement is used to perform the swapping operation, the probability of success is at most \(1/2\). In this way, as the channel length increases so too does the number of repeater nodes and therefore the number of swapping operations that need to be performed. To prevent degradation of entanglement from the entanglement swapping operations, entanglement purification protocols are necessary \cite{bennett1996purification}. 
\begin{figure}
	\centering
	\includegraphics[width=0.95\linewidth]{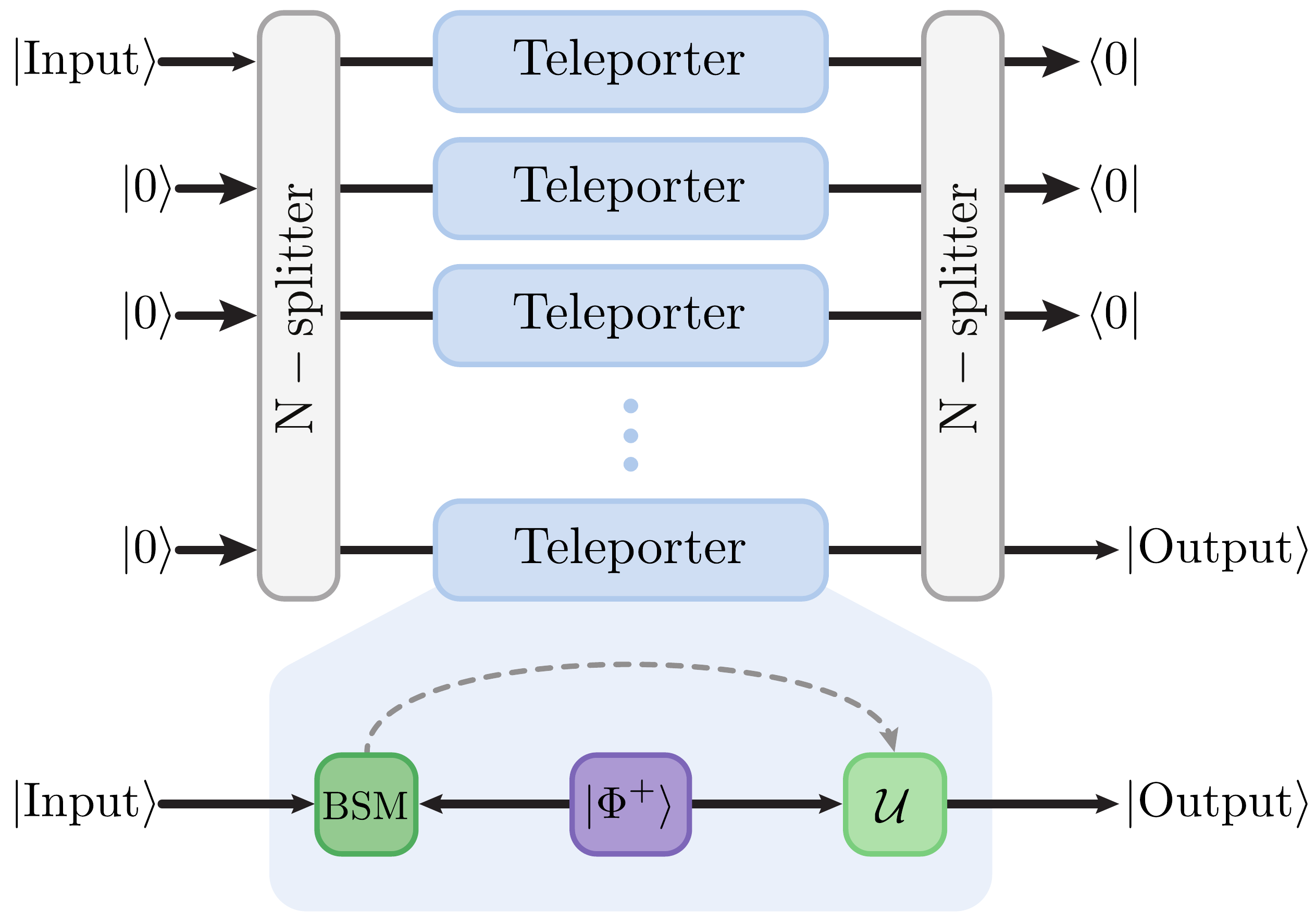}
	\caption{Scheme to teleport CV states using DV entangled resources using the protocol from Ref.~\cite{andersen2013high}. An input CV state is split on an array of beam-splitters with each mode then teleported individually using a qubit teleportation protocol (pictured in blue inset).  As the mean photon number of the input state increases, so does the number of modes. When all qubit teleporters are successful, the modes are recombined on another beam-splitter array.}
	\label{fig:DVTeleporter}
\end{figure} 
Entanglement purification proceeds by distributing two pairs between two repeater nodes (whether these be adjacent or at longer ranges). Within each node, a unitary is applied to the qubit of one pair and the qubit of the second. Following those unitary operations, one of the pairs is measured with the first entangled pair being kept if the measurement results are the same and discarded if the measurement results are different. In this way, two entangled pairs of fidelity \(F\) can result in a single entangled pair of fidelity \cite{bennett1996purification}: 
\begin{equation}
F_{pur}=  \frac{F^2+ \frac{1}{9} \left(1-F\right)^2}{F^2+ \frac{2}{3} \left(1-F\right)+ \frac{5}{9} \left(1-F\right)^2} 
\label{eq:Fpur}
\end{equation}
where the fidelity of the purified pair \(F_{pur}\) is higher than that of the initial two pairs. Entanglement purification is probabilistic however, and the probability of successful purification depends on the fidelity of the initial pairs \cite{bennett1996purification}, 
\begin{equation}
P_{pur} =  \frac{1}{4} \left[F^2 +\frac{2}{3} \left(1-F\right) + \frac{5}{9} \left(1-F\right)^2  \right] .
\end{equation}
where we have included the factor of \(1/4\) that arises from the use of linear optical gates. 
Therefore, while entanglement swapping may be used to distribute entanglement over a long channel using many repeater nodes, the unfortunate consequence of using imperfect pairs means fidelity will degrade for longer channels. Entanglement purification is required at longer distances to circumvent this, the cost being the wait time for the purification to succeed and the extra entangled pairs that are needed. 

In operating the entire repeater, the number of pairs initially distributed will depend on both the number of nodes along the channel and the fidelities of the required final pair and the initial distributed pairs. As an example of how the entire repeater protocol might operate consider Figure~\ref{fig:DVRepeater}. The protocol begins by distributing many different copies of entangled pairs between the nodes. The initial distribution is followed by one round of purification, taking two entangled pairs to a single entangled pair of higher fidelity. A Bell-state measurement is conducted at the second and fourth nodes and after the correction depending on the measurement outcome, entanglement is held between the first and third nodes and the third and fifth nodes. Further rounds of entanglement purification and swapping follow. After all rounds of purification and swapping have succeeded, entanglement is held between both ends of the long channel. We note that with low fidelity initial pairs, many rounds of entanglement purification will be required to produce an end pair of high target fidelity. Additionally for operation with many repeater nodes, the many rounds of entanglement swapping will degrade the fidelity such that entanglement purification will be necessary to achieve an end pair of high target fidelity. Both of these aspects will be important factors governing the rates achievable with DV repeaters discussed later in Section~\ref{sec:rate}.
\begin{figure*}
	\centering
	\subfloat[]{%
		\includegraphics[width=0.6\textwidth]{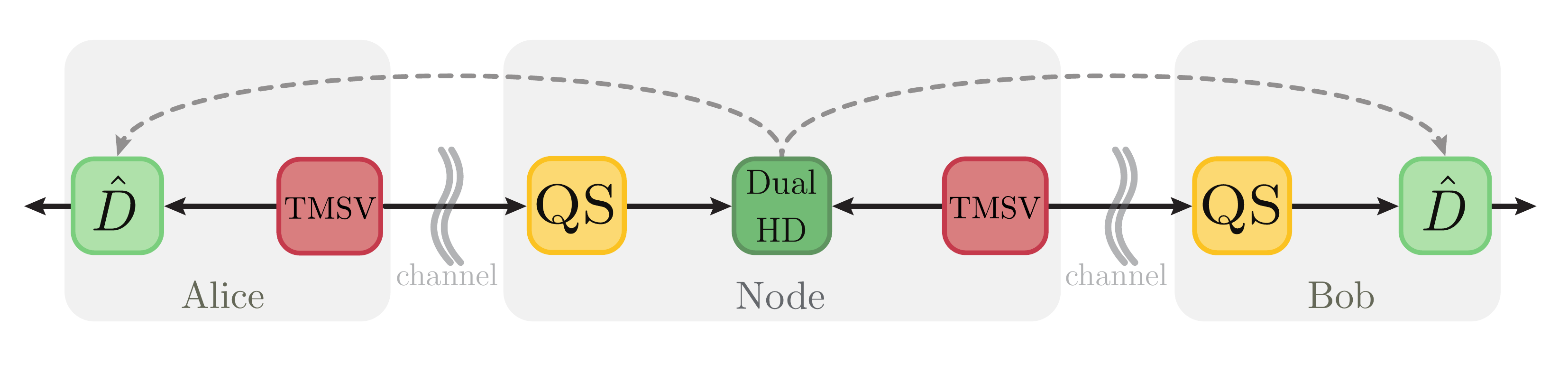} \label{fig:AsymSwap}%
	}
	
	\subfloat[]{%
		\includegraphics[width=0.8\textwidth]{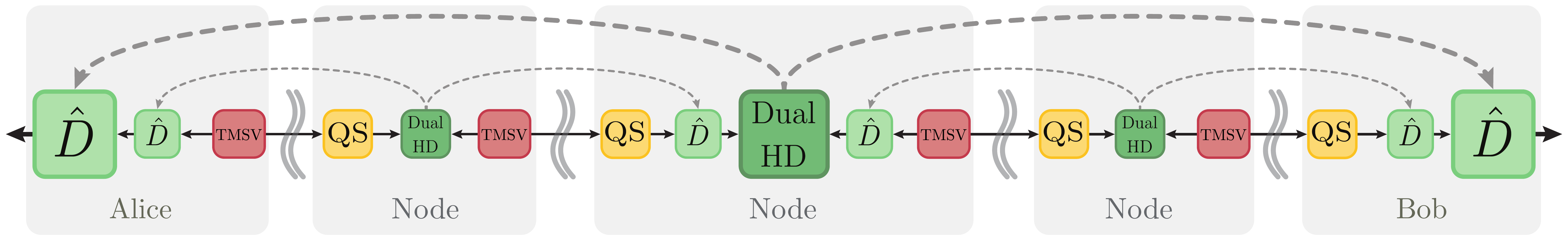}
		\label{fig:nestedAsym}
	}
	\caption{Quantum repeater for continuous variables from Ref.~\cite{dias2020quantuma}. \protect\subref{fig:AsymSwap} Single node implementation with asymmetric distribution of TMSV entangled states. The arm of the entanglement that has passed through the lossy channel is then acted upon by a probabilistic NLA to distill the entanglement. Results in this work focus only on NLA implementations with a single photon quantum scissor, therefore NLAs are depicted here by the yellow boxes labeled `QS'. The two entangled states are then combined in a dual homodyne detection with classical signals sent to both ends of the channel to make displacements accordingly. \protect\subref{fig:nestedAsym} The repeater is scaled up over larger distances by using nested entanglement swapping. Pictured here is the three node implementation.}
	\label{fig:CV}
\end{figure*}
\subsection{Teleporting CV states using DV resources\label{subsec:DV}}
Using the DV repeater protocol outlined in the previous section, we may achieve distribution of discrete variable entangled resource states. By employing a specific teleportation protocol, we may use these entangled resource states to teleport any continuous variable quantum state. This teleportation protocol, conceived by Andersen and Ralph in 2011 \cite{andersen2013high}, is pictured in Figure~\ref{fig:DVTeleporter} and proceeds as follows: an input CV state to be teleported is split on an array of beam splitters (N-splitter) which splits the state evenly among many different modes. The number of modes is dependent on the size (average photon number) of the input state. Each mode is then input into its own discrete teleportation protocol (pictured in the blue inset in Figure~\ref{fig:DVTeleporter}). Here, entangled states are distributed between both ends of the channel. Distributed dual-rail entangled states are then converted to a single rail encoding such that (ideally) $\ket{01}\ket{10}+\ket{10}\ket{01} \to \ket{0}\ket{1} +\ket{1}\ket{0}$. This transformation can be done deterministically using adaptive phase measurements~\cite{ralph2005adaptive}. Then, the sender mixes each of the modes with their qubit of the  Bell state and conducts a Bell-state measurement. Results of the measurement are communicated classically to the receiver who then conducts a unitary operation to the other qubit of the entangled pair. This results in each reduced amplitude mode being teleported individually. After successful teleporation of all the modes, they are then coherently recombined on an N-splitter. When all the ports of the N-splitter register \(\ket{0}\), the output state has been recombined and the teleportation is successful. In our comparison, we restrict ourselves to low energy states and so set the number of modes to 1, significantly simplifying the set-up.
\begin{figure}
	\centering
	\includegraphics[width=0.49\textwidth]{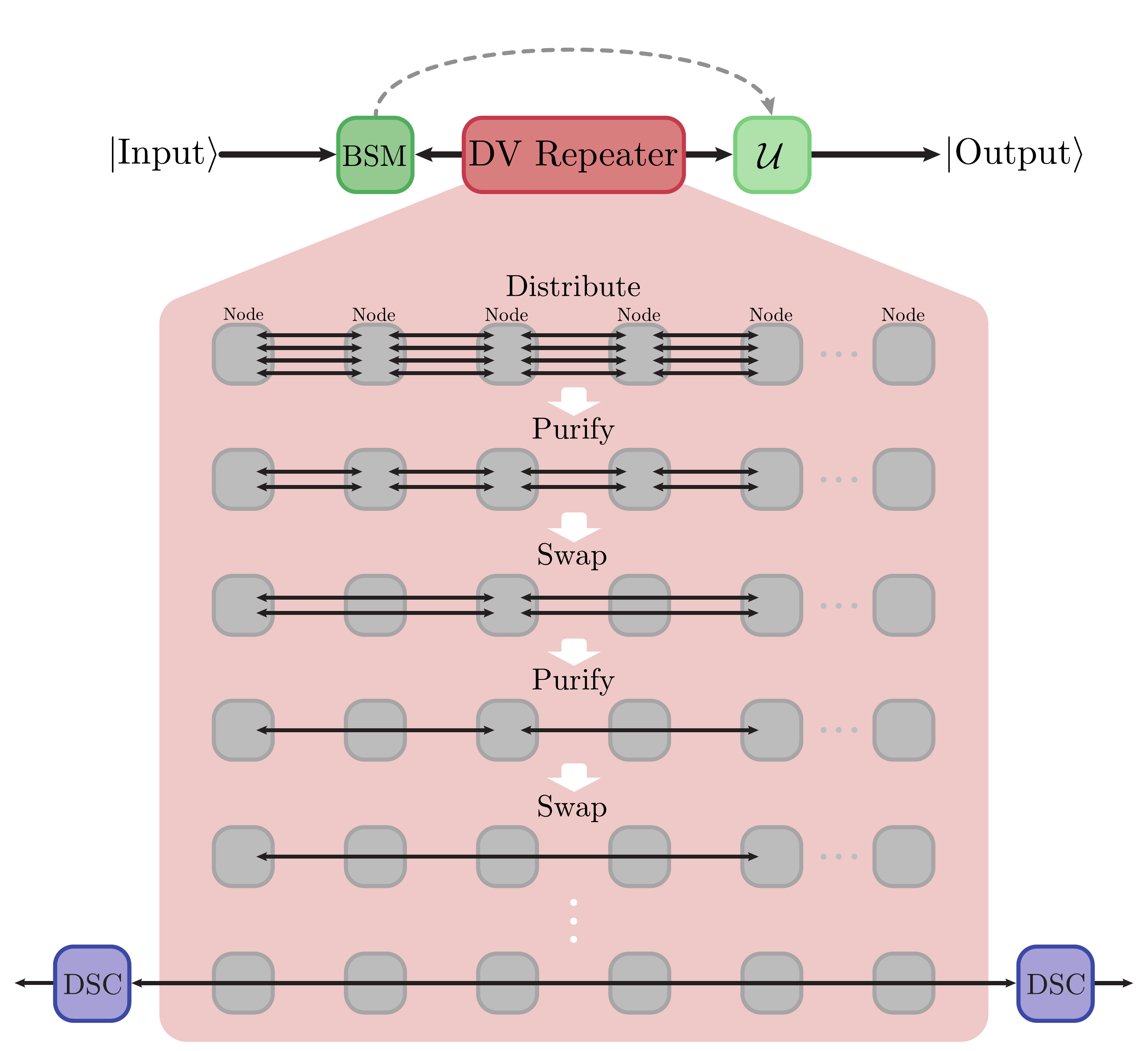} 
	\caption{Distributing CV states using DV repeaters. By employing the discretised teleporation protocol from Ref.~\cite{andersen2013high}, we may distribute CV entangled states over large distances using only DV entangled resources. First-generation DV repeaters may be used to distribute DV entangled pairs over large distances. Once the DV repeater is successful in distributing an entangled pair, the duail-rail entangled pair will need be to converted to a single-rail encoding. The boxes labelled `DSC' represent the dual-rail to single-rail conversion \cite{ralph2005adaptive}. The entangled pair may then be used for qubit teleporation of a CV state provided the input photon number is low. }
	\label{fig:TeleporterRepeater}
\end{figure}

\section{Continuous variable repeaters\label{sec:CV}}
Let us now briefly review the CV repeater protocol from Ref.~\cite{dias2020quantuma}.  Like its discrete variable counterpart, the CV repeater from Ref.~\cite{dias2020quantuma} contains entanglement distribution, entanglement swapping and entanglement distillation. However, each of these elements are  different from the previously described discrete versions so as to be compatible and effective on continuous variable encodings of quantum information. 

This repeater protocol is based on an earlier CV repeater protocol \cite{dias2017quantum}, which itself is based on a protocol to correct CV states against loss errors \cite{ralph2011quantum}. The protocol begins by distributing CV entangled resource states between ends of the channel (or nodes of the repeater). These entangled resources states are the TMSV states given in \eqref{eq:EPR}. Distribution of these states is performed asymmetrically with one arm of the entangled state passing through the lossy channel while the other arm of the entangled state remains in the repeater node.  Entanglement distillation is performed on the arm of the entanglement that has been decohered by loss via the Noiseless Linear Amplifier (NLA) \cite{ralph2009nondeterministic}. We focus on the single-photon ($N=1$) quantum scissor \cite{pegg1998optical} implementation of the NLA. After successful operation of the NLA, the entangled states are combined  and a joint homodyne detection is performed at the repeater  node. Post-selection is used on the measurement results, accepting results that are close to zero. On successful outcomes, the results of this measurement are sent to both ends of the channel and a displacement is performed according to the results of the homodyne detection. 
 
As the total channel distance scales up, more segments of the channel  are required to overcome the drastic effect of loss over very long distances. In this repeater, entanglement is distributed in each segment and then distilled via the NLA and then connected via nested entanglement swapping. Thus, entanglement distillation only occurs directly after entanglement distribution and not at any time after. 
  This is shown in Figure~\ref{fig:CV}\subref{fig:nestedAsym} where three repeater node implementation is pictured. In this setup, two copies of the protocol from Fig.~\ref{fig:CV}\subref{fig:AsymSwap} are used to distribute entanglement between Alice and the middle node and Bob and the middle node. Following this, the modes at the central repeater node are then combined and entanglement swapping is performed via homodyne detection with post-selection. For even longer distances and more segments of the channel, the repeater protocol is scaled up via nesting the entanglement swapping in this way. We note that in this CV repeater, entanglement distillation (via NLA) is probabilistic and due to the use of post-selection, the entanglement swapping is also probabilistic.

\section{Rate comparison\label{sec:rate}}
\begin{figure*}
\centering
\includegraphics[width=0.99\textwidth]{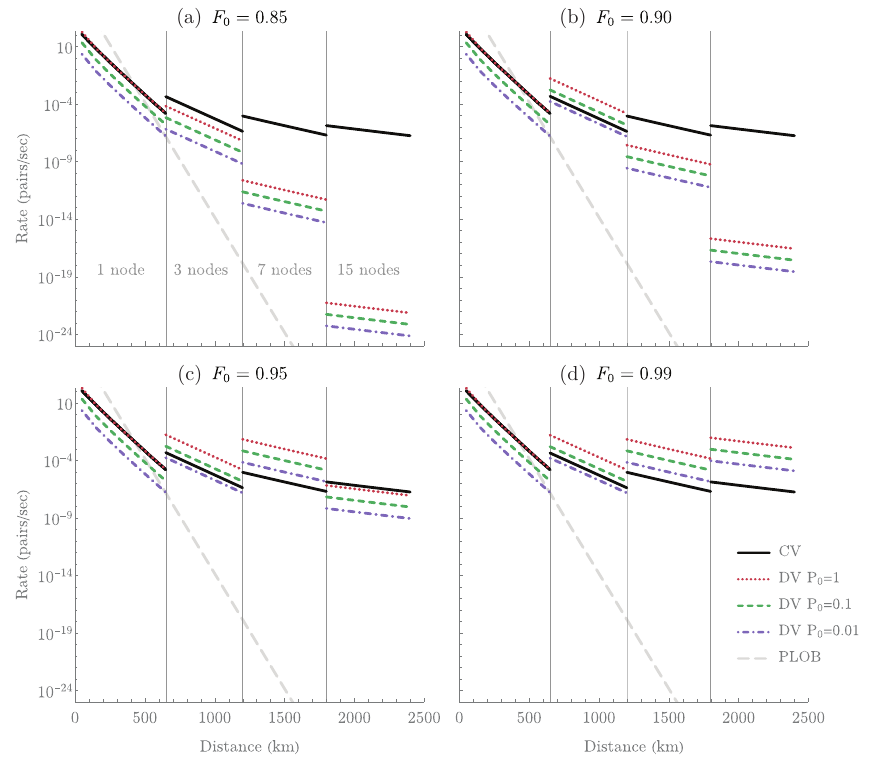}
 \phantomsubfloat{\label{fig:85}} \phantomsubfloat{\label{fig:90}}\phantomsubfloat{\label{fig:95}} \phantomsubfloat{\label{fig:99}}
  \vspace{-2\baselineskip}
  \caption{Entanglement generation rates for CV and DV repeaters. In all figures, the colored lines correspond to DV repeater rates with the red, dotted line showing the outcome assuming deterministic ($P_0=1$) pair production of all required entangled pairs. The green, dashed lines show the rates if this pair production was probabilistic with success probability of $10\%$. The purple, dot-dashed lines show the rates if this pair production was probabilistic with success probability of $1\%$. The black solid lines shows the rates of the CV repeater from Ref.~\cite{dias2020quantuma}. As we wish to compare the rates fairly, we allocate the same number of repeater nodes to each scheme with the number of nodes labeled accordingly. \protect\subref{fig:85} shows the DV rates assuming an initial fidelity of all pairs to be $F_0=0.85$. \protect\subref{fig:90}, \protect\subref{fig:95} and \protect\subref{fig:99} show the same for initial fidelities $F_0=0.90$, $F_0=0.95$ and $F_0=0.99$ respectively. Lower initial fidelities will require more rounds of purification to achieve the target output state, yielding lower repeater rates. The CV (black, solid) lines in all plots are the same. Also shown is the PLOB bound \cite{pirandola2017fundamental} assuming a source rate of 1MHz (gray, dashed line).
}
\label{fig:Results}
\end{figure*}
We have now reviewed all the protocols utilized in our comparison of DV and CV repeater rates. Our goal in this work is to compare the rate of distribution of CV entanglement (rather than QKD key rates) that can be achieved using CV and DV resources. While there are many factors that can affect performance of the repeaters that cannot be directly compared between CV and DV regimes,  our efforts to make the comparison fair will be outlined in this section. Before presenting a detailed discussion we give a succinct summary of our assumptions:
\begin{enumerate}[label=(\roman*)]
\item Photon production, squeezed state production, detection and memory are assumed ideal in both cases.
\item Propagation loss must be corrected by NLA in the CV case. Propagation loss is heralded in the DV case but reduces the probability of success.
\item Operations in the DV system are more complicated and require high quality Bell pairs. We recognize this by ``handicapping" the DV system by limiting the fidelity of the initially distributed Bell pairs.
\end{enumerate} 

Input states to both repeaters will be two-mode Gaussian squeezed vacuum states \eqref{eq:EPR}. Distribution of these states using the DV repeater will follow the protocol pictured in Figure~\ref{fig:TeleporterRepeater}, whereby the DV repeater is used to generate entangled pairs between ends of the channel, and those entangled pairs are then used for teleportation of the CV input state. It worthwhile to note that teleporatation of CV states in this way requires multiple modes if the average photon number is greater than 1. Therefore, this approach will require many DV repeaters running in parallel to achieve distribution of CV states of high mean photon number. 

We are comparing the DV repeater distribution protocol pictured in Fig.~\ref{fig:TeleporterRepeater} to the CV protocol in Fig.~\ref{fig:CV}. Here the CV repeater is used to generate entanglement between both ends of the channel. While the entangled states will be decohered somewhat due to the elements in both repeaters, we are comparing situations where the output of both DV and CV protocols have the same amount of entanglement. The entanglement measure we use is the entanglement of formation (EOF) \cite{bennett1996mixed}. 

While the operations that take place at each node are different in the CV and DV protocols, we approximately allocate the same resources to both repeaters by firstly ensuring we compare the same number of nodes.  Additionally, we make a number of idealized assumptions about both protocols including perfect efficiency photon sources and perfect efficiency detectors as well as ideal memories. We do however have to be careful about how we treat the entangled source used in the DV repeater. Here we consider two situations: the ideal case where a deterministic source is available and a second one built from the pair source (here we assume our entangled state is generated with a either \(10\%\) or \(1\%\) probability). We also only assume linear optics capabilities with these rate comparisons. As an example, this means that each entanglement swapping operation in the DV repeater carries an extra 1/2 factor in the probability of success due to the linear optics construction of the Bell-state measurement. Our comparison also takes into account the time needed for classical communication of successful results and allowing all probabilistic operations to succeed assuming finite resources. This was achieved following the methods in Refs.~\cite{bratzik2013quantum, bernardes2011rate} respectively.

 Given the previously outlined assumptions and restrictions on this comparison, we present in Figure~\ref{fig:Results} the \textit{repeater rate}, in units of entangled pairs per second as a function of total channel distance in kilometers. The results in Fig.~\ref{fig:Results} show the entanglement generation rates for both DV and CV repeaters for varying numbers of repeater nodes along the channel and varying initial infidelities of the DV entangled resource pairs. Here, we have modeled both the DV and CV protocols distributing an entangled state of the same EOF. We emphasize that the repeater rates of entangled pair distribution in Fig.~\ref{fig:Results} are not for pure, maximally entangled pairs, instead each data point produces the same EOF and the rates among the CV and DV schemes may be compared. The EOFs were chosen as to optimise the secret key rate of CV QKD [31]. That is, the CV repeater is working in a high-fidelity and highly-Gaussian regime where the truncation noise due to the first-order NLA is small.

In discussing the results of Fig.~\ref{fig:Results}, let us acknowledge the general behavior of the repeater rate and distributed EOF as a function of the tunable parameters. For the CV scheme, the tunable parameters are the gain of the NLA and the squeezing of the TMSV sources in each node. For results in Fig.~\ref{fig:Results}, the CV repeater parameters have been roughly optimized over gain and squeezing (optimized for secret key rate) and we note for longer distances optimal operation of the CV repeater occurs for lower squeezing (seven and fifteen node results in Fig.~\ref{fig:Results} uses \(\chi=0.11\) and \(\chi=0.08\) respectively). For the DV scheme, there are two tunable parameters governing the final EOF: the target fidelity of the final pair $F_T$ and the squeezing \(\chi\) of the TMSV state input into the discretised teleporter (lower half of Fig.~\ref{fig:RepeaterComparison}). In order to achieve as fair a comparison as manageable, we aim to use TMSV sources of the same strength squeezing in both CV and DV schemes. Then we match the output EOF of both schemes, this sets a certain target fidelity $F_T$ of the final pair in the DV scheme, where the final pair is used for qubit teleportation. In Fig.~\ref{fig:Results}, the one and three node results use the same squeezing for both CV and DV schemes ($\chi=0.32$ and $\chi=0.17$ respectively corresponding to mean photon numbers of $\bar{n}=0.11$ and $\bar{n}=0.03$). However, at seven and fifteen nodes, the CV repeaters use such low squeezing and produce relatively high EOF, the DV repeaters were unable to achieve this same EOF (even with a perfect fidelity final pair \(F_T=1\), see Appendix \ref{app:DV}). For these results at seven and fifteen nodes, the CV squeezing used was \(\chi_{\mathrm{CV}}=0.11\) and \(\chi_{\mathrm{CV}}=0.08\) respectively however the DV squeezing was increased to $\chi_{\mathrm{DV}}=0.3$. While the resource allocation may not be perfectly the same in this specific aspect, we emphasize that all the points in Fig.~\ref{fig:Results} produce entangled output states of the same EOF. The EOF for all results in Fig.~\ref{fig:Results} ranges from 0.04 to 0.11 in units of ebits.

Now that we have established reasonable grounds for a fair comparison, we can discuss the performance of the repeater rate. For the DV schemes, this is heavily impacted by the fidelity of the initial pairs generated between nodes, as low initial fidelities will require more rounds of purification. In Fig.~\ref{fig:Results} we give the repeater rates for various initial fidelities from $F_0=0.85$ in \ref{fig:Results}\subref{fig:85} to $F_0=0.99$ in \ref{fig:Results}\subref{fig:99}. Note that in Fig.~\ref{fig:Results}, the CV rates (black solid lines) are the same in \ref{fig:Results}\subref{fig:85}-\ref{fig:Results}\subref{fig:99} as we are only varying the initial fidelities of the entangled pairs in the DV schemes. At $F_0=0.85$, the most drastic decrease in repeater rate is witnessed with the 15 node results needing a total of 10 rounds of purification resulting in a maximum repeater rate on the order of $10^{-22}$. Conversely, with an initial fidelity of $F_0=0.99$, the DV repeater can achieve the target EOF with no rounds of purification needed at all, thus achieving the maximum rates. Further, the one node results in \ref{fig:Results}\subref{fig:85}-\ref{fig:Results}\subref{fig:99} are all the same, with no purification rounds needed to achieve the target EOF. However, as distance is increased, so too is the need for more repeater nodes and thus more rounds of entanglement swapping which then introduces the need for entanglement purification. In this way, we can observe that while the single node results are the same for \ref{fig:Results}\subref{fig:85}-\ref{fig:Results}\subref{fig:99} ($F_0\geq0.85$), the three node DV results are the same for \ref{fig:Results}\subref{fig:90}-\ref{fig:Results}\subref{fig:99} ($F_0\geq0.90$) and the seven node results are the same for \ref{fig:Results}\subref{fig:95}-\ref{fig:Results}\subref{fig:99} ($F_0\geq0.95$). In all aforementioned cases, the results are the same because no rounds of purification are needed to achieve the target EOF from these initial fidelities.

From the results in Fig.~\ref{fig:Results}, we can see that the single node CV repeater rate overlaps with the single node DV deterministic rate. For higher distances using more repeater nodes, the DV rates are either better or worse than the CV repeater rates depending on the initial fidelity of the distributed pairs and the the source probability. In particular, we can see that all the DV repeater rates are significantly worse than the CV repeater rates for 15 node operation with initial fidelity of $F_0=0.85$ or $F_0=0.90$. However, all DV rates are better than the CV repeater rates for 15 node operation when the initial fidelity is $F_0=0.99$. From this, we can see that access to high fidelity pairs may produce faster entanglement distribution with DV resources and operations, however CV schemes may be more useful when this is not the case.

Additionally, it can be seen in Fig.~\ref{fig:Results}, that while we have restricted the number of repeater nodes to certain fixed distances, this does not represent optimal operation in all cases. For example, in \ref{fig:Results}\subref{fig:95}, the repeater rate for the DV deterministic case (red, dotted line) specifically for seven nodes will achieve faster rates over three node operation at some distances less than 1200km and also faster than fifteen node operation at some distances greater than 1800km. This can be seen by the position of the red, dotted line for seven nodes in \ref{fig:Results}\subref{fig:95}, relative to the red, dotted lines for three and fifteen nodes. In this specific case, optimizing repeater rate would result in seven node use for a wider range of distance beyond what is shown in this plot of Fig.~\ref{fig:Results}\subref{fig:95}. However, for simplicity and ease of comparison between different initial fidelities, we have fixed all the node operation to be within certain distances.

 As the average photon number of one arm of this two-mode squeezed state is less than one, results in Figure~\ref{fig:Results} present the DV teleporter operating on one mode only, even though it could operate with multiple modes (as in Fig.~\ref{fig:DVTeleporter}). Future work in this comparison could benefit from examining distribution of high energy CV states requiring multiple DV modes to distribute. Further, it should be mentioned that the DV scheme is using significantly more Werner states than pair sources in the CV situation. One may want therefore to consider normalized rates but this is left for future work as it is not clear how a fair comparison could be done in such a case. 
\section{Conclusion \label{sec:conc}}
In summary, there is no clear cut answer as to which repeater is more efficient in this highly simplified and idealized comparison. Our simplistic comparison assumes perfect sources and detectors as well as infinite memory coherence times. With single node operation, the CV repeater rate overlaps with the DV (deterministic source) repeater rates. At larger distances and more nodes, the more efficient scheme depends on the source probability and fidelity of the DV entangled pair sources. If many rounds of purification are needed in the DV scheme, the rates achievable via CV schemes will easily surpass the DV rates. Therefore, if one is only able to establish lower fidelity links between repeater nodes, employing DV strategies may not be more efficient at all.

We also emphasize our comparison here is limited to low energy input states, and low EOF distribution ($\mathcal{E}<0.11$ ebits). Attempting to distribute higher energy CV states via DV resources would require multiple DV repeater modes running in parallel which would undoubtedly slow rates. In the CV protocol we consider in this work, this would require NLAs implemented with higher order quantum scissors ($N>1$) \cite{guanzon2022ideal}. This would also yield a significant reduction in entanglement distribution rates as the success probability of the NLA decreases exponentially with order of quantum scissors $N$. We expect the aforementioned effects would likely also be the case for distribution of high EOF states. Nevertheless, our result highlights how DV resources may be effectively utilized to distribute CV states. While our comparison is limited to early repeater protocols of both CV and DV encodings, it would also be of considerable interest how optimisations of the protocols we consider in this work affect the rate of entanglement distribution. In particular,  we expect significant performance increases as the second and third generation CV repeater schemes are developed. 

\section{Acknowledgements}
This research was funded by the Australian Research Council Centre of Excellence for Quantum Computation and Communication Technology (Project No. CE110001027)  and the  JSPS KAKENHI Grant No. 21H04880.

\appendix
\section{Rate comparison}
In this appendix, we give further details of the calculations necessary to achieve Fig.~\ref{fig:Results} of the main text, specifically how we match the amount of entanglement in the output states of CV and DV repeater protocols. After matching the amount of entanglement in the output states, we have calculated the repeater rates for the successful operation of the schemes including all necessary probabilistic operations with finite resources and classical communication times following the methods presented in Refs.~\cite{bratzik2013quantum, bernardes2011rate}. 
\subsection*{Continuous variable}
Firstly, the CV repeater results shown are the upper bound results given in Ref.~\cite{dias2020quantuma}. These results were produced in a high-fidelity and highly Gaussian regime where the truncation noise due to the single photon NLA is small. For each data point, the first and second moments of the output state yield the covariance matrix of an equivalent Gaussian state from which we can calculate the Gaussian entanglement of formation (GEOF) \cite{wolf2004gaussian, marian2008entanglement} of the distributed entangled state. Recall that while the output states of the CV repeater are not exactly Gaussian, as noted earlier, we are operating the repeater in a highly Gaussian regime.  Also note that for two mode Gaussian states, the exact entanglement of formation coincides with the GEOF \cite{marian2008entanglement,akbarikourbolagh2015entanglement}. 
\subsection*{Discrete variable \label{app:DV}}
\begin{figure}
	\centering
	\includegraphics[width=0.8\columnwidth]{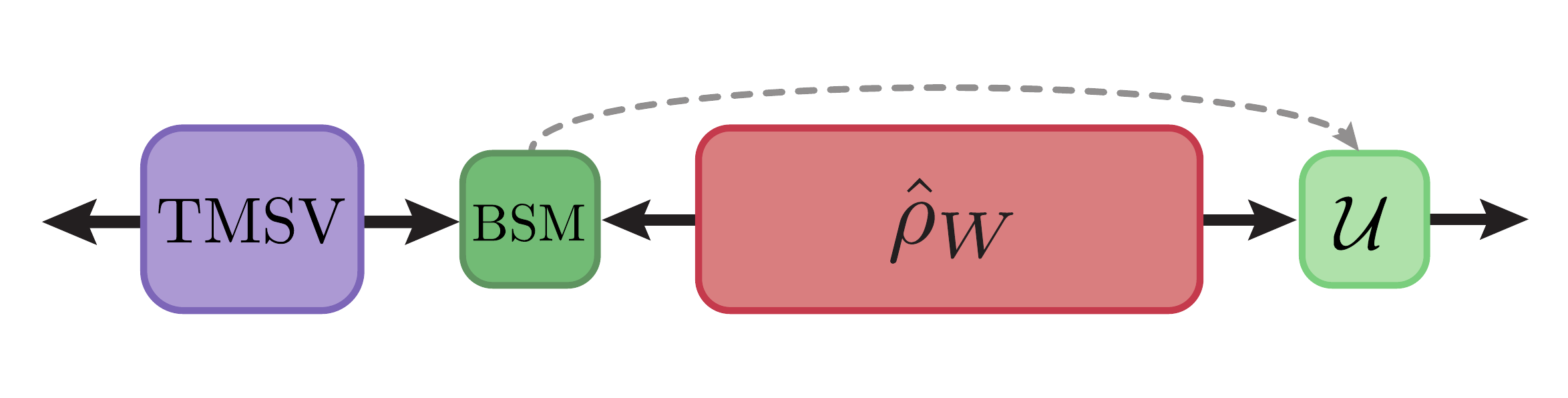}
	\caption{DV teleportation protocol with a TMSV state input. After distribution of a noisy entangled pair (modelled by a Werner state $\rho_W$  \eqref{eq:Werner}) is completed via the DV repeater, that pair is used for CV teleportation via the protocol from Ref.~\cite{andersen2013high}. When this teleportation protocol is implemented with a single mode, we have the scheme pictured above. }
	\label{fig:DVtele}
\end{figure}
\begin{figure}
	\centering
	\includegraphics[width=0.8\columnwidth]{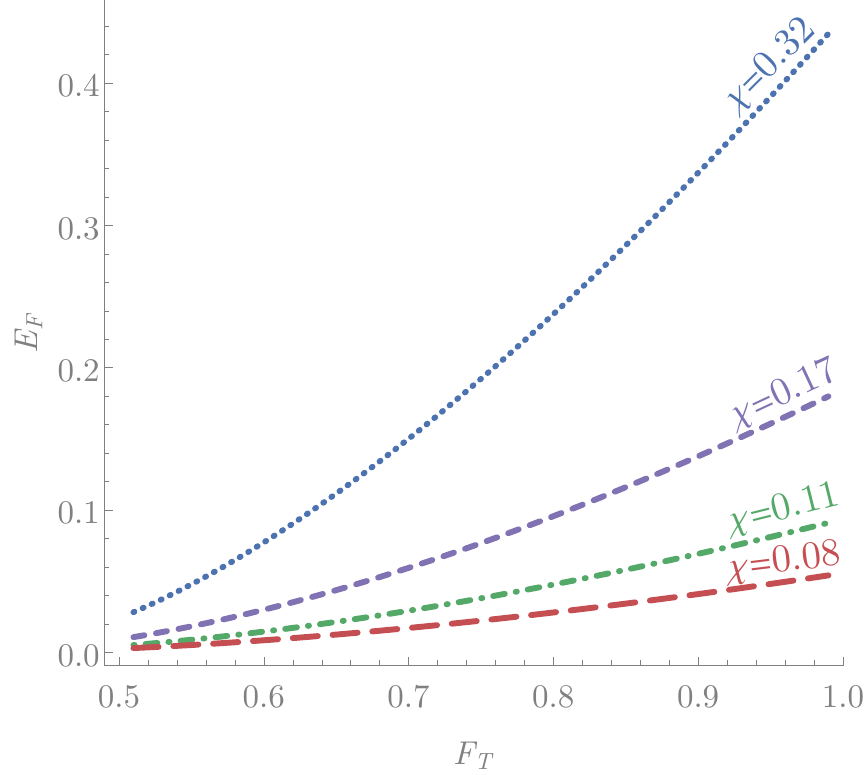}
	\caption{Entanglement of formation of the output state \eqref{eq:teleOut} of the DV teleportation protocol. The entanglement of formation is depends on both the squeezing of the input state $\chi$ and the fidelity of the Werner pair used as the resource state for teleportation. The four values of $\chi$ shown here of $\chi=0.32$, $0.17$, $0.11$ and $0.08$ were used to generate the CV results in Fig.~\ref{fig:Results} for 1, 3, 7 and 15 nodes respectively. }
	\label{fig:eof}
\end{figure}
In the DV case, teleporation of one arm of the TMSV state \eqref{eq:EPR}  is performed using the CV teleportation protocol from Ref.~\cite{andersen2013high} with a Werner state \eqref{eq:Werner} of target fidelity $F_T$ as the entangled teleportation resource. We consider operating the teleportation protocol from Ref.~\cite{andersen2013high} with a single mode, which is possible as $\chi$ is kept small. The protocol is pictured in Fig.~\ref{fig:DVtele}. The input state to be teleported and the entangled resource state have the combined form:

\begin{align}
{{\rho}}_{1234}=	&\ket{\chi}_{12}\bra{\chi}_{12} \otimes{{\rho}_w}_{34}
\end{align}
where $\ket{\chi}_{12}$ is given by~\eqref{eq:EPR} and ${{\rho}_w}_{34}$ is given by~\eqref{eq:Werner}. Recall that while we consider distributing dual-rail entangled states (see Sec.~\ref{sec:DV}), they are converted to a single rail encoding before teleportation. 

A Bell-state measurement on modes $2$ and $3$ followed by the corresponding unitary transformation on mode $4$ gives the following output state:
\begin{equation}
	{{\rho}} _{14}=\frac{1}{\chi^2+1}\begin{bmatrix}
\frac{2F_T+1}{3} & 0 & 0& \chi \left(\frac{4F-1}{3}\right)\\
0 & 2\chi^2 \left(\frac{1-F_T}{3} \right) & 0 & 0 \\
0 & 0 &   \frac{2-2F_T}{3} & 0 \\
\chi \left(\frac{4F_T-1}{3} \right) & 0 &0 & \chi^2 \left(\frac{2F_T+1}{3} \right)
	\end{bmatrix}
\label{eq:teleOut}
\end{equation}
in the \(\{\ket{00}, \ket{01}, \ket{10}, \ket{11}\}\) basis. Note that the output \eqref{eq:teleOut} is an entangled state of two qubits because of the single mode teleporter. Additionally, since the output is an entangled state of two qubits, we can easily compute its exact entanglement of formation \cite{wootters1998entanglement}. 

The EOF of this output state of DV protocol output is limited by the fidelity of the final pair $F_T$ and the strength of the squeezing of the TMSV input state. In Fig.~\ref{fig:eof}, we show the EOF of this protocol as a function of the fidelity of the distributed entangled resource state $\hat{\rho}_W$. The four lines given in Fig.~\ref{fig:eof} are for $\chi=0.32$, $0.17$, $0.11$ and $0.08$ and correspond to the values used for the CV results in Fig.~\ref{fig:Results} for 1, 3, 5 and 15 nodes respectively. It can be seen that the EOF for low values of $\chi$ is limited even with perfect target fidelity pairs ($F_T=1$). This is particularly evident in the curves for $\chi=0.11$ and $0.08$ which are the values used for CV operation with 7 and 15 nodes respectively, and this is the reason the DV schemes were allowed access to squeezing of $\chi=0.3$ for the seven and fifteen node results - it was necessary in order to match EOF of the output to the CV scheme.

We need to remember that the initial EOF of the input TMSV states with low squeezing is also low, however, the CV repeater is able to output states of higher EOF than the EOF of the TMSV input state. In the DV scheme,  while we are able to distill the DV pairs to arbitrary high fidelity, once we use that pair for teleportation of one arm of an entangled state, the output EOF of the teleported entanglement can never surpass the EOF of the input state. In this comparison, by matching the EOF of the DV output to the CV output, this sets a certain target fidelity \(F_T\) that is required for teleportation. The number of rounds of purification needed to reach this target fidelity thus varies according to the target fidelity itself as well as the initial fidelty of the pairs distributed between nodes.

\end{document}